\documentclass{elsart}

\usepackage{natbib}

\usepackage{epsfig}

\usepackage{amssymb}

\begin{document}

\begin{frontmatter}



\title{Improved orbit predictions using two-line elements}


\author{Creon Levit\corauthref{cor}}
\address{NASA Ames Research Center, Moffett Field, MS202-3, CA 94035, USA}
\corauth[cor]{Corresponding author}
\ead{creon.levit@nasa.gov}

\author{William Marshall}
\address{NASA Ames Research Center and Universities Space Research Association, Moffett Field, MS202-3, CA 94035, USA}

\begin{abstract}

The density of orbital space debris constitutes an increasing environmental challenge. There are three ways to alleviate the problem: debris mitigation, debris removal and collision avoidance. This paper addresses collision avoidance, by describing a method that contributes to achieving a requisite increase in orbit prediction accuracy. Batch least-squares differential correction is applied to the publicly available two-line element (TLE) catalog of space objects. Using a high-precision numerical propagator, we fit an orbit to state vectors derived from successive TLEs.  We then propagate the fitted orbit further forward in time. These predictions are compared to precision ephemeris data derived from the International Laser Ranging Service (ILRS) for several satellites, including objects in the congested sun-synchronous orbital region. The method leads to a predicted range error that increases at a typical rate of 100 meters per day, approximately a 10-fold improvement over TLE's propagated with their associated analytic propagator (SGP4). Corresponding improvements for debris trajectories could potentially provide initial conjunction analysis sufficiently accurate for an operationally viable collision avoidance system.  

We discuss additional optimization and the computational requirements for applying all-on-all conjunction analysis to the whole TLE catalog, present and near future. Finally, we outline a scheme for debris-debris collision avoidance that may become practicable given these developments. 

\end{abstract}


\end{frontmatter}

\parindent=0.5 cm


\section{Introduction}
Collisions in orbit pose a threat to spacecraft, to astronauts and to the global commons of near-earth space.  Several collisions have already occurred between spacecraft and debris, while the Iridium 33/Cosmos 2251 collision of January 2009 represented the first documented satellite-satellite collision. Unfortunately, the cumulative number of collisions thus far is consistent with the prescient predictions of a runaway chain reaction \citep{kessler1978}.

Reducing debris from current and future space missions (debris mitigation) is an important measure.  However, even the unrealistic case of ``no new launches" is inadequate to curb runaway growth \citep{liou2008}. There are two remaining ways to alleviate the problem: debris removal and collision avoidance. For the latter, one needs prediction precision to be sufficiently accurate so as not to imply an unwieldy number of collision avoidance maneuvers. In addition, one needs a means of doing debris-debris collision avoidance since applying collision avoidance only to the subset of conjunctions involving maneuverable spacecraft does not suffice to curb the growth. This paper describes a method to increase the orbit prediction accuracy based on publicly available TLEs. 

Many satellite owner operations have inadequate (if any) processes for conjunction assessment  and collision avoidance, since they would have to screen their asset(s) against all other space objects. The only source of knowledge at their disposal for the majority of other objects is the publicly available two-line element (TLE) sets.  But  predictions based on TLEs using the associated analytic propagator (SGP4) are not sufficiently accurate to warrant maneuvering to avoid potential collisions: they imply an unacceptably large number of potential collisions per space object, each of which has very low probability. The problem is similar for debris-debris conjunctions except then both objects, not just one, are subject to these imprecisions. 

To address this TLE/SGP4 accuracy problem, we investigated several methods to improve the propagation errors for non-maneuvering orbital objects whilst using only TLEs as input data.  The following research target was set: ``increase the predictive accuracy for orbital objects, using only historical TLE data, such that it enables operational conjunction assessment for collision avoidance''.   

In the following sections, we describe one approach to this target, assess its accuracy, and discuss the requirements for extension to the entire space object catalog. Finally, we propose a new method of debris-debris collision avoidance enabled by long-term high-accuracy conjunction assessment.

\section{Method: TLE orbit fitting and propagation}

ESA uses TLEs from the publicly available catalog to initially screen their sun-synchronous orbit (SSO) spacecraft ERS2 and ENVISAT for conjunctions \citep{flohrer2009}. Telemetry from their operational spacecraft provide precision orbital ephemeredes (POEs) for those spacecraft.  For screening against all other potentially conjuncting objects, only TLEs are used.  \citet{flohrer2008} describe a method to estimate error covariances of TLEs in order to quantify collision probability assessments. Their method provided inspiration for the present work.  Here we extend and adapt their approach: based solely on the object's historical TLEs, we improve the accuracy of the object's predicted position, as opposed to quantifying the accuracy of the object's SGP4 propagation errors.

Our method is essentially to use TLE data as ``pseudo-observations'' and to fit an orbit to these pseudo-observations using a high-precision special perturbations propagator and traditional batch least-squares differential correction.  The fitted orbit is then propagated into the future using the same high-precision orbit propagator.  The prediction accuracy is assessed by comparison with precision orbital ephemeris (POE) data from the International Laser Ranging Service \citep{pearlman2002}.  

For each object we wish to analyze we choose a time window with two sections: a fitting period and a subsequent prediction period.  The fitting period is initially set to ten days, typical of the period over which U.S. Space Surveillance Network observations are fitted when generating TLEs for LEO objects \citep{danielson2000}. Section $4$ details a more principled approach to determining the fitting period. The prediction period is 30 days.

For a given object, all TLEs with epochs within its entire window are obtained. However, only TLEs with epochs within the fitting period are used for the fitting process; those in the prediction period are only used for validation and test. 

Using all TLEs from the fitting period and interpolating using SGP4, we generate a series of order $100$ state vectors (``pseudo-observations'') equally spaced in time within the fitting period.  We initialize the differential correction with a state vector derived from the first TLE in the fitting period, and then propagate that using a high precision propagator. We apply batch least-squares differential correction to minimize the RMS error in the radial, in track and cross track (RIC) components of relative positions between the fitted orbit and the pseudo-observations. The trajectory obtained from the converged differential corrector is our fit. 

The high precision propagator incorporates a $60\times 60$ EGM2008 gravity field including solid Earth tides, point masses for the solar and lunar gravity fields and a NRL-MSISE $1990$ atmospheric drag model.   We assume constant values for solar F$10.7$ and Ap equal to their averages over the fitting period.  Solar radiation pressure is modelled using a simple the biconic approximation with Earth and Moon as eclipsing bodies.    

\section{Results: Prediction Accuracy}

\begin{figure}[ht]
\label{figure1}
\begin{center}
\includegraphics*[width=14cm,angle=0]{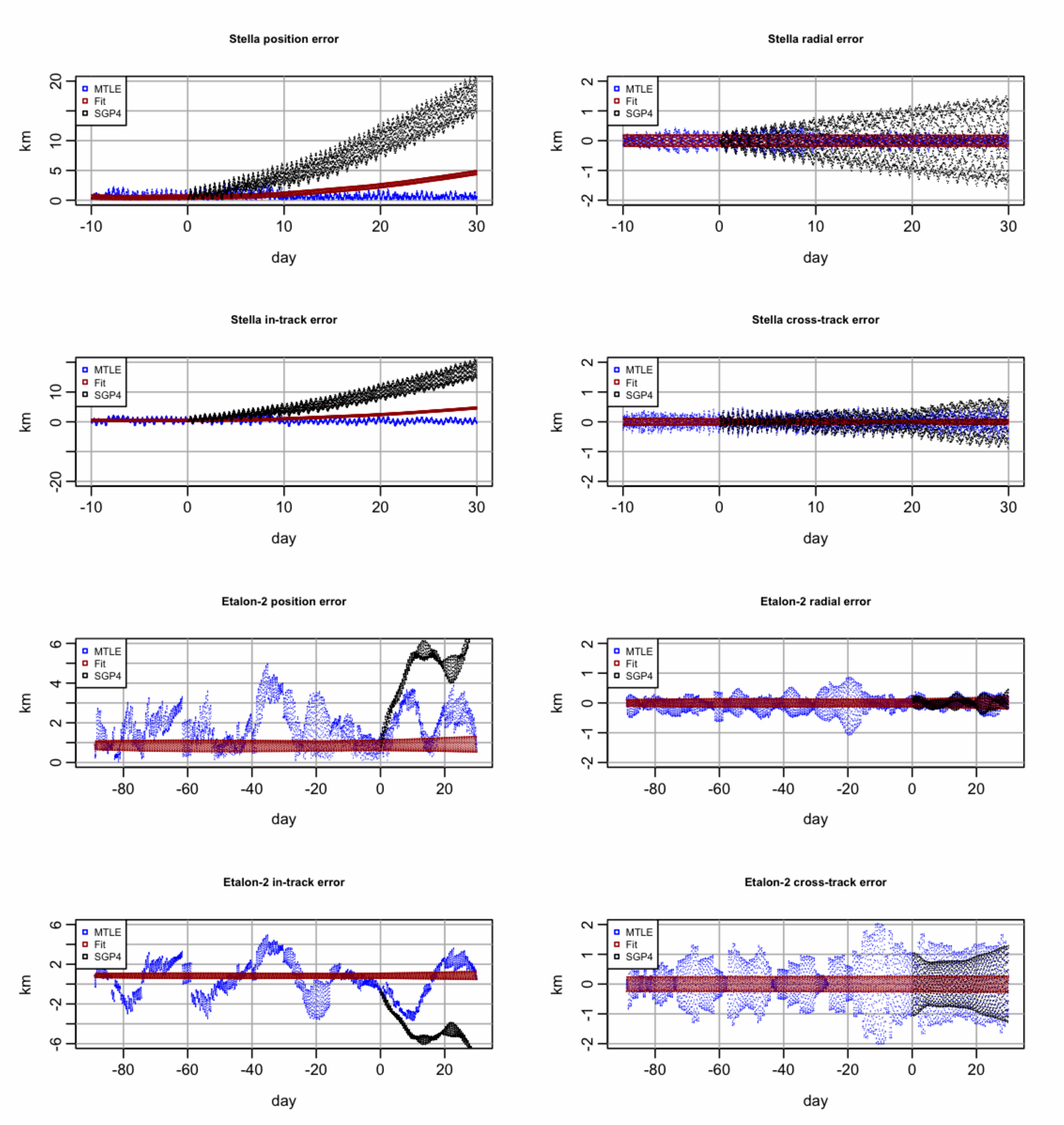}
\end{center}
\caption{
Typical position errors with respect to POE (``truth'') using our method (red), and using SGP4 (black) for Stella (upper panels) and Etalon-2  (lower panels). The prediction period is $0<t\le30$ days.  The fitting period is $-10 \le t \le 0$ days for Stella and $-90 \le t \le 0$  days for Etalon-$2$. All fitting uses only publicly available TLEs. Truth data are only used for plotting. Selection of the length of the fitting period is discussed in Section 4. Also shown are the updated TLEs (no prediction) in blue.
 }
\end{figure}

We applied our method to four non-manoeuvring spacecraft for which POE data were readily available \citep{ries2009}:  Stella, Starlette, Ajisai and Etalon-$2$. The summary of basic orbital and physical properties for these spacecraft are shown in Table $1$. They were picked in part for their range of orbital altitudes.   

\begin{table}[h]
\caption{Key Parameters of Geodesic Spacecraft Used}
\begin{tabular}{ l c c c c c c}
\hline
Satellite&Perigee&Eccentricity&Period&Inclination&Mass&Diameter\\
&(km)&&(min)&(deg)&(kg)&(m)\\
\hline
Stella & 800 & 0.0206 & 98.6 & 101 & 48 & 0.24 \\
Starlette & 812 & 0.0206 & 104 & 49.8 & 47 & 0.24 \\
Ajisai & 1490 & 0.0010 & 116 & 50.0 & 685 & 2.15 \\
Etalon 2 & 19120 & 0.0007 & 675 & 65.5 & 1415 & 1.29 \\
\hline
\end{tabular}
\label{table2}
\end{table}

Figure $1$ shows examples for two of these spacecraft -- the highest and lowest altitude -- comparing our method's predictions to those of SGP4. The TLEs for these objects exhibit quite different behaviours.  Nevertheless, in both cases the TLEs propagated with SGP4 (black) depart rapidly from the true position of the satellite whereas our numerically fitted and numerically propagated orbits (red) maintain higher accuracy, particularly over the long term.  

\begin{figure}[ht]
\label{figure2}
\begin{center}
\includegraphics*[width=14cm,angle=0]{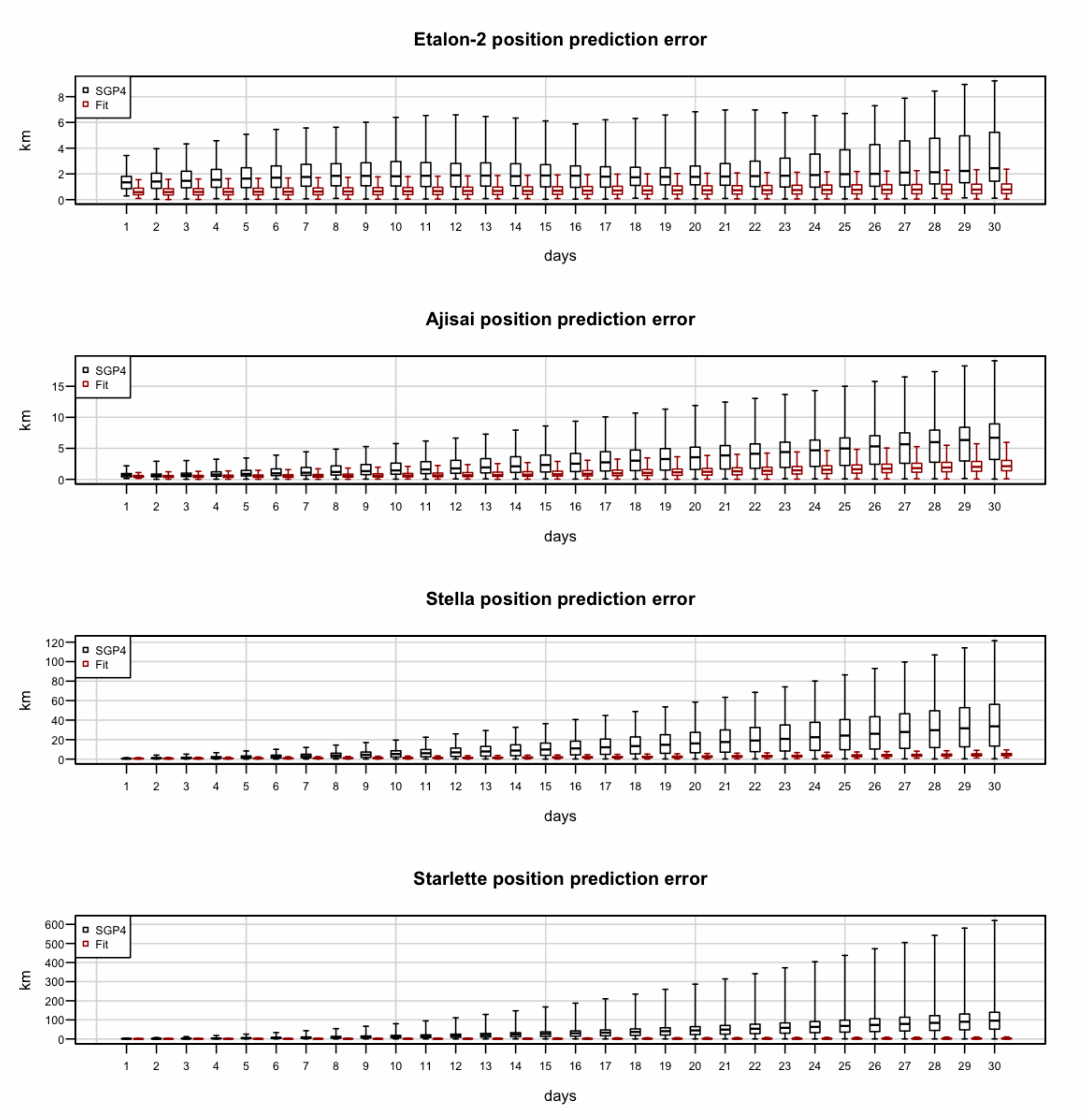}
\end{center}
\caption{ 
``Box and whisker'' plots of range errors for predictions using our fit method (red), and using SGP4 (black) for four satellites.  Each panel summarizes $50$ runs corresponding to $50$ random starting dates in $2004$.  The $x$-axis is prediction time in days. The $y$-axis is prediction error in km. The lower and upper whiskers extending from each box bound the minimum and maximum prediction errors respectively. The lower and upper box edges bound the $25^{th}$ and $75^{th}$ percentiles, respectively. The glyphs inside the boxes mark the median prediction error. All errors are plotted with respect to ``truth'' (i.e. POEs obtained from the ILRS). Truth data are only used for plotting.  All fitting uses only publicly available TLEs.}
\end{figure}

\begin{figure}[ht]
\label{figure3}
\begin{center}
\includegraphics*[width=14cm,angle=0]{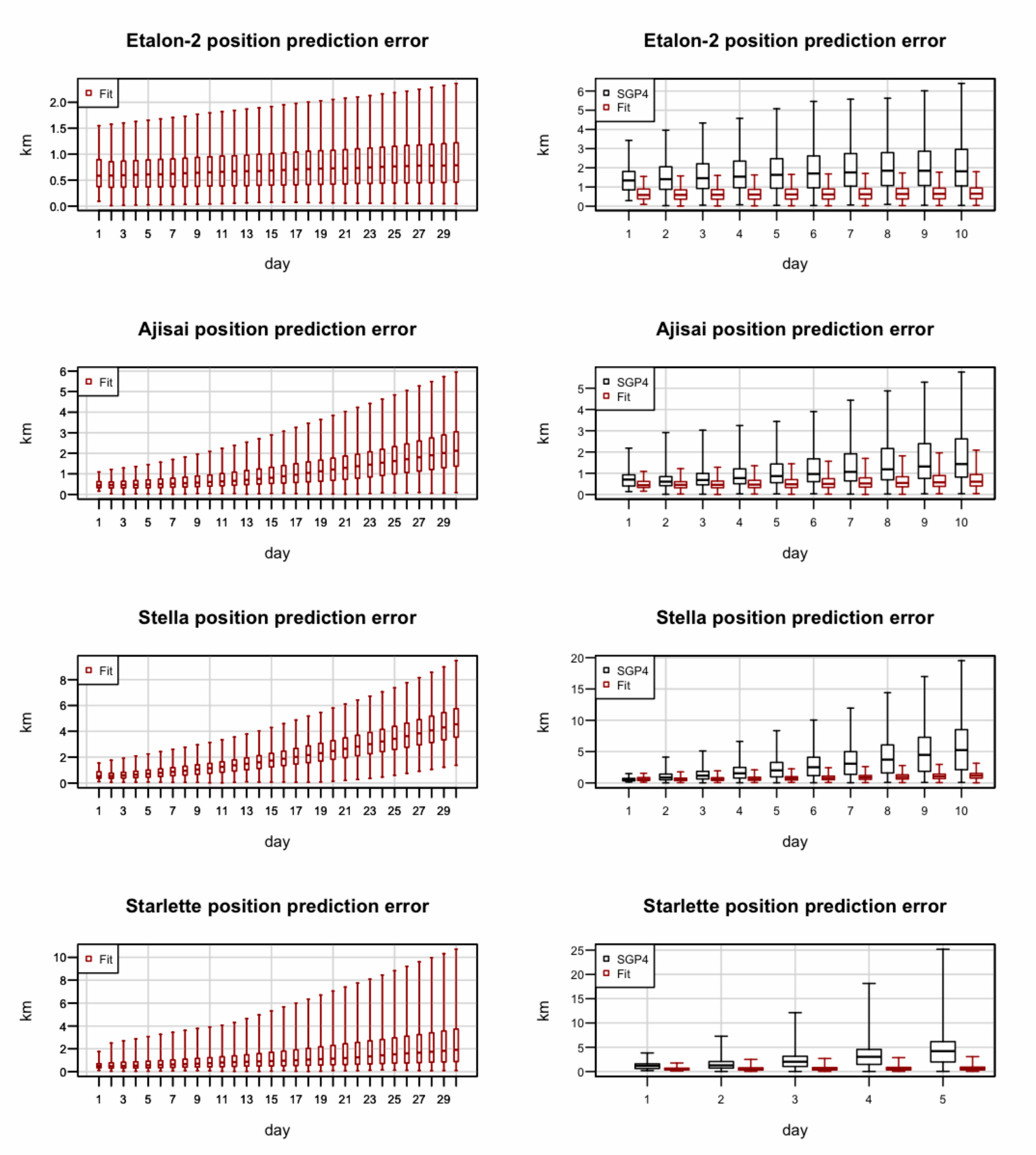}
\end{center}
\caption{
Additional summaries of range errors for predictions using our method (red), and using SGP4 (black) for four satellites. On the LHS, just our method is plotted for a full $30$ days and on the RHS both our method and the SGP4 propagated TLEs are shown but just for the first $10$ days ($5$ in the case of Starlette since the TLE was diverging rapidly).  Each box and whisker plot summarizes $50$ runs corresponding to $50$ random starting dates in 1994.  The $x$-axis is prediction time in days. The $y$-axis is prediction error in km.  All errors are plotted with respect to ``truth'' (i.e. POE obtained from the ILRS). Truth data are only used for plotting.  All fitting uses only publicly available TLEs.
}
\end{figure}

Precision orbit ephemeredes (``truth data'') define the $x$-axis for plotting, but truth data are not used in the fitting process. Nor are they used for tuning hyperparameters (e.g. length of fitting interval, see Section $4$).  Improvements in prediction accuracy are typically seen in all three axis (RIC), especially more than day into the future. For predictions of less than one day the errors appear to be dominated by TLE bias (with respect to truth) which our method cannot remove.

For each of the four satellites, we performed $50$ runs of our method using different starting dates distributed uniformly in $2004$. Summary statistics of these runs using ``box-and-whisker'' plots appear in Figures $2$ and $3$. The TLE+SGP4 errors grow more rapidly in every case.   The typical prediction errors for our method are $3$km at $30$ days out, corresponding to $\sim 100$ m/day prediction error growth. The worst case is Stella where the median error grows $<150$ m/day and the best is Etalon at $<30$ m/day. The maximum error for our method over all four satellites is $<300$ m/day. These compare to typical growth of $1,500$ m/day for TLE's propagated with SGP4 for these objects. The ratio of median prediction error using our method vs. SGP4 at $30$ days range from $3$ for Etalon-$2$ to $50$ for Starlette, and averages $15$. 

Thus for these satellites, our method exhibits approximately one order of magnitude improvement in prediction accuracy over TLEs propagated with SGP4. Since the improvement is in all three directions (RIC) the resultant decrease in position covariance ellipsoid volume is likely to result in $2$ to $3$ orders of magnitude reduction in false positive conjunctions.

Since the TLEs for these objects have an instantaneous range bias from truth of $0.8 \pm 0.3$ km the error growth of approximately $1.5$ km/day means that only after approximately one day can one detect the benefits of our method.  Our fitting method has a similar initial range bias to TLEs but error growth of only $100$ m/day. These errors are similar to those published in the open literature evaluating the high accuracy special perturbations catalog(s) maintained by the US Space Command \citep{neal1997,coffey1998,boers2000}. Table 2 compares the accuracy of various tracking data and prediction methods. 

\begin{table}[h]
\caption{Accuracies of Different Sensors and Prediction Methodologies.}
\begin{tabular}{l c c r}
\hline
Method&Prediction&Accuracy&Accuracy\\
&(days)&(m)&(m/day)\\
\hline
Laser Ranging (ILRS ``truth")\footnotemark[1]  & 0 & 0.1& N.A.\\
Fence (raw direction cosines)\footnotemark[2]  & 0 & 10& N.A.\\
High Accuracy catalog + SP\footnotemark[3]  & 10 & 500-2000 & 50-200\\
TLEs + SGP4\footnotemark[4] & 10 & 1,000-30,000 & 100-3,000\\
TLEs + new scheme & 10 & 500-2,000 & 50-200 \\
\hline
\end{tabular}
\label{table1}
\end{table}
\footnotetext[1]{\citet{pearlman2002}}
\footnotetext[2]{\citet{hayden1962,gilbreath1997,schumacher2001}}
\footnotetext[3]{\citet{neal1997,coffey1998,boers2000}}
\footnotetext[4]{\citet{boyce2004, chan2001, flohrer2009, kelso2007, muldoon2008, snow1999, wang2009}}

\section{Fitting Debris objects}

\begin{figure}[ht]
\label{figure4}
\begin{center}
\includegraphics*[width=10cm,angle=0]{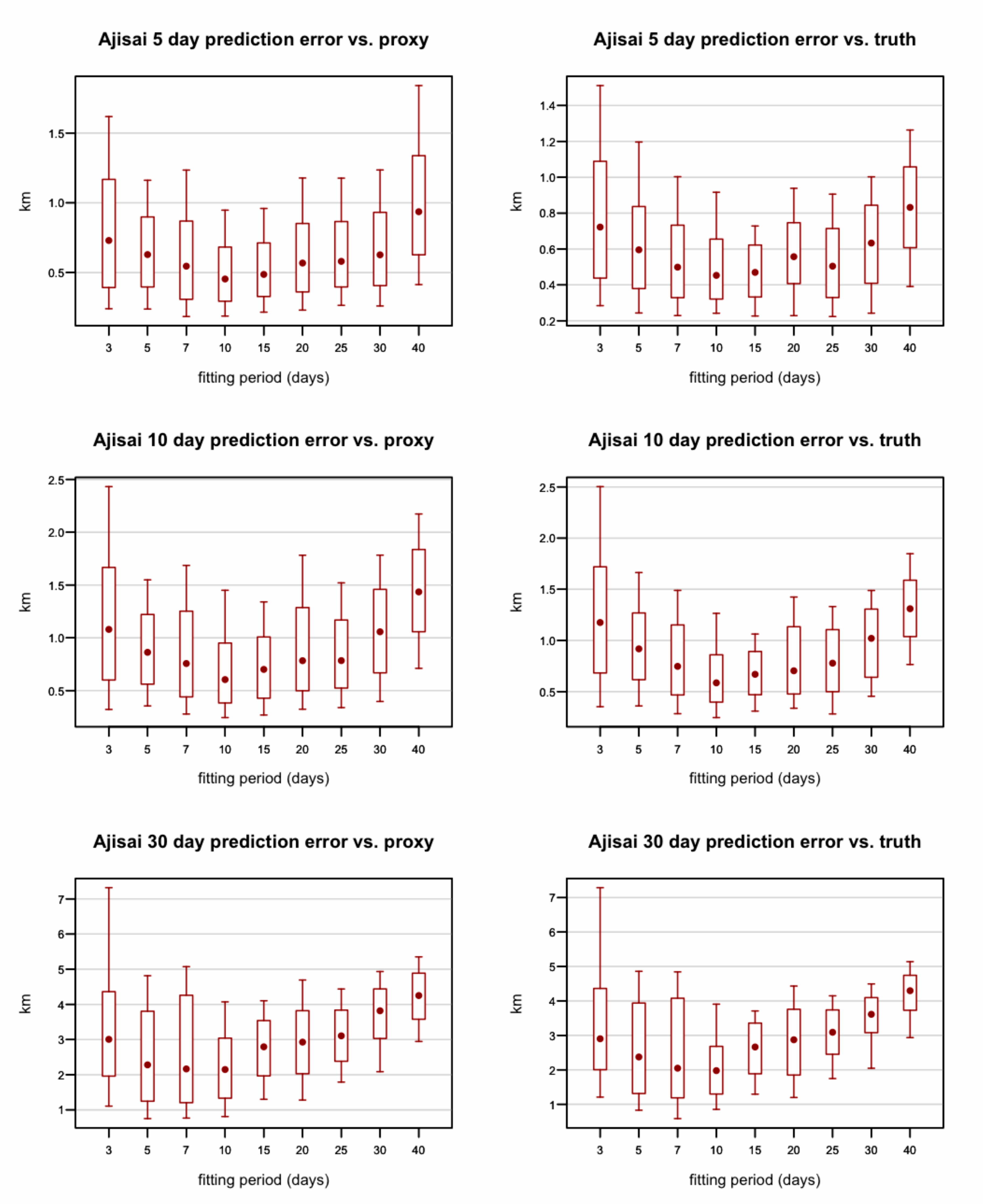}
\end{center}
\caption{
TLEs from beyond the fitting period can be used instead of precision ephemeris (POE) data when optimizing the length of the fitting period.  Left: prediction error with respect to future TLEs (```proxy'') vs. fitting period.  Right: prediction error with respect to POE data (``truth'') vs. fitting period.  Each box and whisker plot summarises $50$ runs corresponding to $50$ random starting dates in $1994$.   
}
\end{figure}

The majority of tracked space objects (i.e debris) do not have precision orbital ephemerides -- they are not tracked by the ILRS and they transmit nothing.  Many are, however, tracked by the United States Space Surveillance Network (SSN).  Unfortunately, the only publicly available source of data derived from these SSN observations are, at present, the TLEs accessible at space-track.org (and archived e.g. at celestrak.com).

While there is currently no publicly available source of high-accuracy trajectory for debris objects, we believe the method described above could be applied to refine the predictions of debris objects' future positions based only upon comparatively inaccurate TLEs, facilitating conjunction assessment and collision avoidance.   However, there are several issues which complicate the application of our procedure to debris:

The first set of issues arise when TLEs are the {\em only} source of data - we have no ``truth'' in order to perform validation and test.  This is solved by the use of TLEs from beyond the fitting period as a proxy for truth data.  

Our initial manual experiments suggested that different fitting periods were optimal for different orbit categories. These results are consistent with \citet{alfriend2002,danielson2000}. We made a more principled investigation, utilizing cross-validation, to optimize the length of the fitting interval on a per-object basis using only TLEs for both fitting and validation.  Figure 4 shows how a fitting period of $10$ days was determined to be optimal for satellite Ajisai. The important point shown in the figure is that a $10$ day fitting period is optimal both when validating predictions with truth data {\em and} when validating predictions with future TLEs.  And the $10$ day fitting period is optimal whether predicting $5$, $10$ or $30$ days into the future.  Thus it appears ``future'' TLEs can be used as a proxy for truth data when optimizing the length of the fitting period on a per object basis. 

The second set of issues is that for debris objects one must also solve for area to mass ratio (drag) and validate this as well using only future TLEs. We are currently performing these experiments.

\section{Outlook}

\subsection{Methodological Improvements}
Further accuracy improvements could come from a principled analysis of the number of fitting points and their weighting as a function of location within the fitting period, the removal of outlier data (i.e. filtering) and other techniques from statistical orbit determination, treating TLEs as observations. Also, recent work \citet{legendre2008, muldoon2008} suggests that there may be simple global transformations of TLE data that can improve TLE/SGP$4$ accuracy.  These systematic corrections might also be further refined on the basis of orbital parameters, resulting in a look-up-table used to correct TLEs as a function of their orbital regimes.  

In addition, the scheme could be tested on actual known conjunctions in order to analyse its efficiency for conjunction assessment. These issues will be the subject of a future paper. 

\subsection{Computational Scaling}
As an initial assessment of the computational requirements needed to perform conjunction analysis for all space objects in the publicly available catalog, a simple conjunction analysis system was parallelized\footnote
{
This work was performed by Chris Henze of the NASA Advanced Supercomputing Division at NASA Ames Research Center.  Details will be forthcoming in a future publication.
}
on the Pleiades supercomputer at NASA Ames Research Center. Using approximately $200$ CPU cores, checking all objects against all objects for conjunctions $7$ days into the future takes about a minute.  The conjunctions involving spacecraft precisely match the results from SOCRATES \citep{socrates}.  The same system takes about $40$ minutes to perform 7 days of all-on-all conjunction assessment of a ``simulated S-band fence'' catalog  (obtained from the NASA Orbital Debris Office) containing ``pseudo-TLEs'' for approximately $2.5$ million objects of size $\ge 2$cm. This demonstrates the feasibility of doing all-on-all conjunction assessment on the present and expected near-future catalogs. Scaling the fitting method discussed above to the future catalog will be the subject of a later paper.

\subsection{A New Scheme for Debris-Debris collision avoidance}

An along track $\Delta V$ of $1$ cm/s provides of order 1km displacement per day. Since error growth for our method is $~100$m per day, such small manoeuvres might suffice for collision avoidance. For manoeuvres this small one could use radiation pressure to impart the necessary momentum. To verify the conjecture, we increased the flux of radiation by a factor of 10x the solar radiation constant for 10 minutes on a typical debris object (area to mass ratio $0.1$ m$^2$/kg) in SSO. We then propagated that for a further 5 days. The displacement was $>0.1$ km/day along track compared to a reference object that was not so irradiated. Two cumulative such events, if one ensured appropriate geometry, would thus lead to the necessary along-track displacement.\footnote{This idea was first conceived in discussions with Dr. R\"{u}diger Jehn from ESA, July 2009.}  

A factor of 10x the solar rational constant could conceivably be applied from the ground. For example, a 10kW class laser attached to a 1m class telescope with appropriate tracking, could, (considering diffraction limits only) provide the necessary radiation pressure for a debris object of area $<$1m$^2$ in SSO. One would need to perform a detailed engineering analysis of the feasibility of such a device, particularly the viability of maintaining small beam divergence and tracking. 

Debris-debris collision avoidance could eliminate the need for an active debris removal program. The results of \citet{liou2009} imply, to first order, that performing debris-debris collision avoidance for 5 additional carefully selected objects per annum would curb debris field growth. Furthermore, any additional debris-debris collision avoidance would start reducing the net debris density, eventually to below a critical level, because debris creating collisions are being nulled whereas atmospheric drag continues to cause objects re-entry. Thus this method could provide not just a stop gap measure, but a permanent solution.

\section{Conclusions}

This fitting and propagation method, based solely on TLEs, could potentially provide initial conjunction analysis sufficiently accurate for an operationally viable collision avoidance system. If similar improvements are possible for debris objects, as we expect, then it could be used for all-on-all conjunction assessment.  However, to ensure that the scheme works in an operational setting, it would be necessary to have additional data: the so-called ``uncorrelated objects'' (a.k.a. ``analysts set') which account for approximately 30 percent of potential conjunctions in LEO \citep{Neuman08}. Further, we show that all-on-all conjunction assessment is possible with moderate computer infrastructure, even with the large increase in size of the catalog of tracked objects that is expected in the next few years.  

Finally, given accurate predictions, we claim {\em debris-debris collision avoidance may be possible} by externally inducing small manoeuvres using radiation pressure from $10$ kW class power density ground-based lasers. If feasible, this could negate the need for a large scale  debris removal program.



\begin{thebibliography}{}

\bibitem[Alfriend et al.(2002)]{alfriend2002}
K.~Alfriend, S.~Paik, V.~Boikov, Z.~Khutorovsky, \& A.~Testov.
Comparison of the Russian \& US Algorithms for Catalog Maintenance for Geosynchronous Satellites. Adv. Astronaut. Sci., 112(2):1157--1176, 2002.

\bibitem[Boers et al.(2000)]{boers2000}
J.~Boers, S.~Coffey, W.~Barnds, D.~Johns, M.~Davis, \& J.~Seago.
Accuracy assessment of the naval space command special perturbations cataloging system.
Spaceflight Mechanics 2000, volume 105 of Adv. Astronaut. Sci., pages 1291--1304, 2000.

\bibitem[Boyce(2004)]{boyce2004}
W.~H. Boyce.
Examination of NORAD TLE accuracy using the iridium constellation.
Spaceflight Mechanics 2004, volume 119 of Adv. Astronaut. Sci., pages 2133--2142, 2004.

\bibitem[Chan \& Navarro(2001)]{chan2001}
J.~Chan \& D.~Navarro.
Comparison of NORAD two-line elements with INTELSAT orbital elements.
Proceedings of the Third European Conference on Space Debris, volume SP-473, pages 771--780, 2001.
  
\bibitem[Coffey et al.(1998)]{coffey1998}
S.~Coffey, H.~Neal, C.~Visel, \& P.~Conolty.
Demonstration of a special-perturbations-based catalog in the naval space command system.
Spaceflight mechanics 1998, volume~99 of  Adv. Astronaut. Sci., pages 227--248, 1998.

\bibitem[Danielson et al.(2000)]{danielson2000}
D.~Danielson, D.~Canright, D.~Perini, \& P.~Schumacher.
The naval space command automatic differential correction process.
Astrodynamics 1999, volume 103 of Adv. Astronaut. Sci., pages 991--1008, 2000.

\bibitem[Flohrer, Krag, \& Klinkrad(2009)]{flohrer2009}
T.~Flohrer, H.~Krag, \& H.~Klinkrad.
ESA's process for the identification \& assessment of high-risk conjunction events.
Adv. Space Res., 44(3):355--363, 2009.

\bibitem[Flohrer, Krag, \& Klinkrad(2008)]{flohrer2008}
T.~Flohrer, H.~Krag, \& H.~Klinkrad.
Assessment \& categorization of TLE orbit errors for the US SSN catalog.
In  Advanced Maui Optical \& Space Surveillance Technologies Conference, page~513--524, 2008.

 \bibitem[Gilbreath(1997)]{gilbreath1997}
G.~Gilbreath.
Calibrating the naval space surveillance fence using satellite laser ranging.
Astrodynamics 1997, volume~97 of Adv. Astronaut. Sci., pages~403--416, 1997.

\bibitem[Hayden(1962)]{hayden1962}
L.~Hayden.
Optical calibration of the US Naval Space Surveillance System.
Technical Report NRL 5741, U.S. Naval Research Laboratory, 1962.
  
\bibitem[Kelso \& Alfano(2005)]{socrates}
T. S. Kelso \& S.~Alfano.
Satellite orbital conjunction reports assessing threatening encounters in space (SOCRATES).
Proceedings of the 15th AAS/AIAA Space Flight Mechanics Conference. Paper AAS 05-124, 2005.

\bibitem[Kelso(2007)]{kelso2007}
T.~Kelso.
Validation of SGP4 \& IS-GPS-200 against GPS precision ephemerides (AAS 07-127).
In 17th AAS/AIAA Space Flight Mechanics Conference, volume 127 of  Adv. Astronaut. Sci., page 427, 2007.

\bibitem[Kessler \& Cour-Palais(1978)]{kessler1978}
D.~Kessler and B.~Cour-Palais.
Collision frequency of artificial satellites: The creation of a debris belt.
J. of Geophys. Res., 83(A6):2637--2646, 1978.

\bibitem[Legendre et al.(2008)]{legendre2008}
P.~Legendre, R.~Garmier, G.~Prat, B.~Revelin, and S.~Delavault.
Improvement of the Two-Line Element Accuracy Assessment Based on a Mixture of Gaussian Laws
Adv. Astronaut. Sci., 129(3):2189--2207, 2008.

\bibitem[Liou \& Johnson(2008)]{liou2008} 
J.-C. Liou \& N.~Johnson.
 Instability of the present LEO satellite populations.
 Adv. Space Res., 41:1046--1053, 2008.

\bibitem[Liou \& Johnson(2009)]{liou2009}
J.-C. Liou \& N.~Johnson.
A sensitivity study of the effectiveness of active debris removal in leo.
Acta Astronaut., 64:236--243, 2009.

\bibitem[Muldoon \& Elkaim(2008)]{muldoon2008}
A.~Muldoon \& G.~Elkaim.
Improved orbit estimation using GPS measurements for conjunction analysis.
In Inst. of Navigation Global Navigation Satellite Systems Meeting, pages~786--794, 2008.
  
\bibitem[Neal, Coffey, \& Knowles(1997)]{neal1997}
H.~Neal, S.~Coffey, \& S.~Knowles.
Maintaining the space object catalog with special perturbations.
Astrodynamics 1997, Adv. Astronaut. Sci., pages 1349--1360, 1997.

\bibitem[Neuman(2008)]{Neuman08}
L.~Newman. NASA Goddard Space Flight Center.
 Personal communication, 2009.
 
\bibitem[Pearlman, Degnan, \& Bosworth(2002)]{pearlman2002}
M.~Pearlman, J.~Degnan, \& J.~Bosworth.
The International Laser Ranging Service.
 Adv. Space Res., 30(2):135--143, 2002.

\bibitem[Ries(2009)]{ries2009}
J.~Ries. Precision Orbital Ephemerides (POEs) for satellites Stella, Starlette, Ajisai \& Etalon-2 were obtained from John Ries at the Center for Space Research, University of Texas, Austin TX.

\bibitem[Schumacher et al.(2001)]{schumacher2001}
P.~Schumacher, G.~Gilbreath, M.~Davis, \& E.~Lydick.
Precision of satellite laser ranging calibration of the naval space surveillance system.
J. Guid. Control Dynam., 24(5):925--932, 2001.

\bibitem[Snow \& Kaya(1999)]{snow1999}
D.~Snow \& D.~Kaya.
Element set prediction accuracy assessment.
In Astrodynamics 1999, volume 103 of  Adv. Astronaut. Sci., pages 1937--1958, 1999.

\bibitem[Wang, Liu, \& Zhang(2009)]{wang2009}
R.~Wang, J.~Liu, \& Q.~Zhang.
Propagation errors analysis of TLE data.
Adv. Space Res., 43(7):1065--1069, 2009.



\end{thebibliography}

\end{document}